\documentclass[aps,prl,twocolumn,preprintnumbers,amsmath,amssymb,superscriptaddress,floatfix]{revtex4}

\usepackage{graphicx}
\usepackage{dcolumn}
\usepackage{amsmath}
\usepackage{amssymb}
\usepackage{color}
\usepackage{multirow}
\usepackage{url}
\usepackage{booktabs}
\usepackage{tabularx}

\newcommand\hly{\bgroup\markoverwith
  {\textcolor{yellow}{\rule[-.5ex]{.1pt}{2.5ex}}}\ULon}

\newcommand{\mSR}{\textmu SR}

\newcommand{\refsubfig}[2]{\hyperref[#1]{\ref*{#1}#2}}

\usepackage[utf8]{inputenc}
\usepackage{siunitx} %\SI
\usepackage{verbatim} %\begin{comment}
\usepackage[normalem]{ulem}
\usepackage{placeins} %%\FloatBarrier
\usepackage{float} % [H]
\usepackage[hidelinks]{hyperref} %Hyperref
\hypersetup{
colorlinks=true,    %true,%Switch Print/Online (false, true)
linkcolor=blue,
citecolor=blue,
urlcolor=blue,
pdfborder={0 0 0},
bookmarksopen=true,
bookmarksopenlevel=2,
pdfpagemode=UseOutlines,
pdfstartview={Fit},
pdftitle={},
pdfauthor={Z. Salman},
pdfsubject={Advanced muon-spin spectroscopy},
pdfkeywords={},
pdfhighlight= /I
}%end of hypersetup

\begin{document}
%\preprint{Preprint (\today), please keep it confidential.}
\title {Advanced muon-spin spectroscopy with high lateral resolution using Si-pixel detectors}

\author{Lukas Mandok}
\affiliation{Physics Institute, Heidelberg University, Im Neuenheimer Feld 226, 69120 Heidelberg, Germany}
\author{Pascal Isenring}
\affiliation{PSI Center for Neutron and Muon Sciences, 5232 Villigen PSI, Switzerland}
\affiliation{Physik Institut, University of Zürich, Winterthurerstrasse 190, CH-8057 Zürich}
\author{Heiko Augustin}
\affiliation{Physics Institute, Heidelberg University, Im Neuenheimer Feld 226, 69120 Heidelberg, Germany}
\author{Marius Köppel}
\affiliation{Institute for Particle Physics and Astrophysics, ETH Zürich, CH-8093 Zürich, Switzerland}
\author{Jonas A. Krieger}
\affiliation{PSI Center for Neutron and Muon Sciences, 5232 Villigen PSI, Switzerland}
\author{Hubertus Luetkens}
\affiliation{PSI Center for Neutron and Muon Sciences, 5232 Villigen PSI, Switzerland}
\author{Thomas Prokscha}
\affiliation{PSI Center for Neutron and Muon Sciences, 5232 Villigen PSI, Switzerland}
\author{Thomas Rudzki}
\affiliation{Physics Institute, Heidelberg University, Im Neuenheimer Feld 226, 69120 Heidelberg, Germany}
\author{André Schöning}
\affiliation{Physics Institute, Heidelberg University, Im Neuenheimer Feld 226, 69120 Heidelberg, Germany}
\author{Zaher~Salman}
\email[Correspondnig author: ]{zaher.salman@psi.ch}
\affiliation{PSI Center for Neutron and Muon Sciences, 5232 Villigen PSI, Switzerland}

\date{\today}

\begin{abstract}
Muon-spin spectroscopy at continuous sources has stagnated at a stopped muons rate of \SI{\sim40}{kHz} for the last few decades.
The major limiting factor is the requirement of a single muon in the sample during the typical \SI{10}{\micro s} data gate window.
To overcome this limit and to be able to perform muon-spin relaxation (\mSR) measurements on millimeter-sized samples, one can use vertex reconstruction methods to construct \mSR\ spectra. This is now possible thanks to the availability of very thin monolithic Si-pixel chips, which offer minimal particle scattering and high count rate. Here we present results from a Si-pixel based spectrometer that utilizes vertex reconstruction schemes for the incoming muons and emitted positrons. With this spectrometer we were able to obtain a first vertex reconstructed \mSR\ (VR-\mSR) spectrum. The unique capabilities and benefits of such a spectrometer are discussed.
\end{abstract}

\maketitle

\section{Introduction}
Measuring sub-millimeter samples using \mSR\ is a dream for material scientists, in particular, in the field of novel quantum materials which are difficult to produce in large quantities. In addition, data collection rates at continuous muon sources, which offer high time resolution measurements, have stagnated at a stopped muons rate of \SI{40}{kHz} since the development of the technique. In this paper, we have developed a prototype muon spin relaxation/rotation (\mSR) spectrometer, which breaks these limitations, ushering in a new era for muon-spin spectroscopy and its applications. With recent developments in silicon pixel detector chips used in particle physics experiments, such a concept becomes possible for the first time, benefiting from an excellent single hit spatial resolution of \SI{23}{\micro m} and a combined timing resolution better than \SI{15}{\nano s}~\cite{heikoPhd}.

%Using Si-pixel detectors for \mSR\ measurements is poised to achieve several advancements, including:
%\begin{itemize}
%    \item \textbf{High-Resolution Mapping}: Ability to obtain \textit{magnetic and electronic properties} of materials with \textit{sub-millimeter lateral resolution}, enabling detailed spatial analysis.
%    \item \textbf{Parallel Sample Analysis}: Capability to measure \textit{multiple samples simultaneously}, facilitating \textit{rapid and efficient} exploration of the phase space within a family of materials.
%    \item \textbf{Dynamic Measurements}: Support for \textit{pump-probe and transient experiments}, leveraging a \textit{high data collection rate} to capture time-sensitive phenomena.
%    \item \textbf{Extended Measurement Timeframes}: Reduced uncorrelated background noise, enabling \textit{longer measurement time windows} (beyond the current limit of \SI{10}{\micro\second}) while maintaining both \textit{high time resolution} and \textit{data acquisition rates}.
%\end{itemize}

Other potential breakthroughs for the \mSR\ technique~\cite{himb} include (i) the ability to obtain magnetic and electronic properties of materials on the sub-millimeter lateral resolution, (ii) measure multiple samples in parallel allowing, e.g., fast and efficient exploration of the phase space of a family of materials, (iii) perform pump-probe and transient measurements benefiting from the high data collection rate and measurements of \mSR\ signals with lower uncorrelated background, and (iv) allowing longer measurement time windows beyond the current \SI{10}{\micro s} limit, without sacrificing the high time resolution or data rate.

Here, we demonstrate the feasibility of utilizing the MuPix11 \cite{mupix10,mupix,mupix1,mupix2,mupix3,mupix4,mupix5,mupix6} chip -- a thin high-voltage monolithic active pixel Sensor (HV-MAPS)~\cite{Peric2007NIMA} -- to construct a prototype \mSR\ spectrometer. This prototype is designed to operate efficiently in a high-rate continuous muon beam at the Paul Scherrer Institute (PSI).
The technology has been developed in the context of the Mu3e experiment~\cite{mu3e}, currently under construction at PSI.
This work demonstrates two key performance aspects of the design: First, acquiring \mSR\ spectra with sub-millimeter lateral resolution from the sample region. This will allow the separation of regions (on the scale of 100s of microns) with different magnetic properties in the samples as well as the measurement of multiple samples in parallel.
Second, the superb spatial resolution will be used to identify multiple muons implanted in the samples at the same time, enabling an increase in the incoming muon rate by more than tenfold, and therefore, make more efficient use of the highly oversubscribed \mSR\ spectrometers/beamlines at \mSR\ user facilities.

\section{Experimental}
\begin{figure}[h]
\centering
\includegraphics[width=\columnwidth]{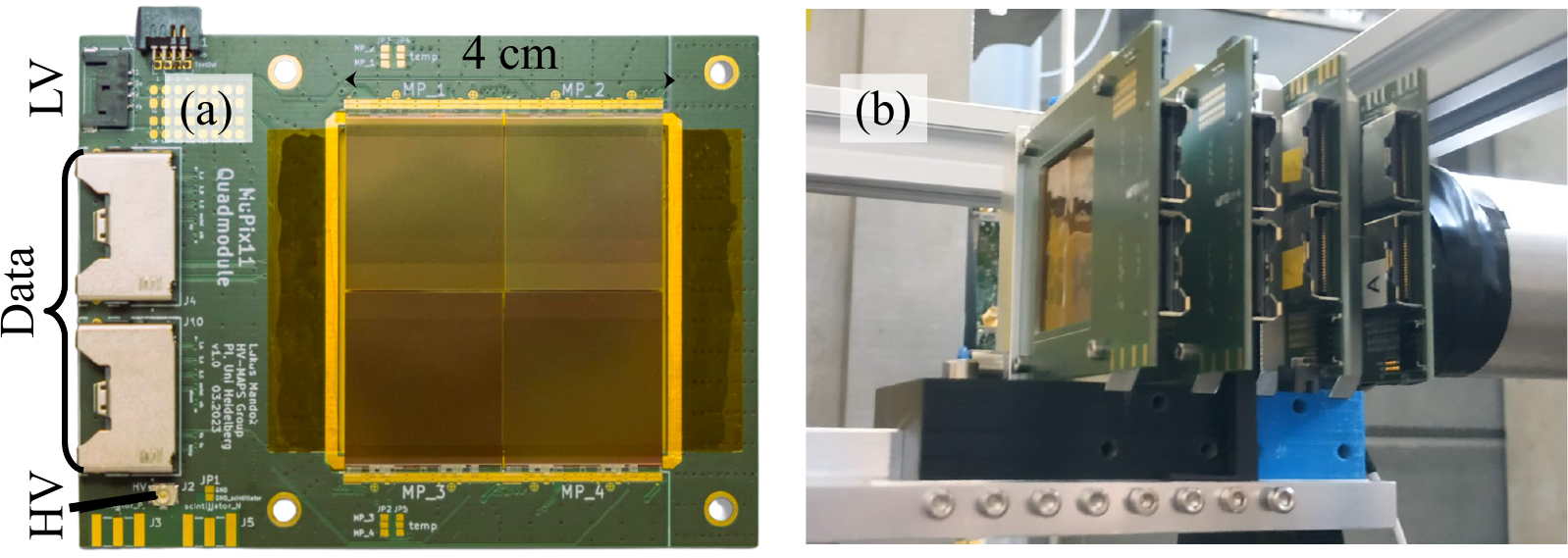}
\caption{\label{spectrometer} (a) A photo of a Quad hosting four Si-pixel chips. (b) The arrangement of the Quads in the prototype spectrometer at the $\pi$E3 beamline at PSI with two upstream, two downstream, with the investigated sample in the middle.}
\end{figure}

\begin{figure}[h]   \centerline{\includegraphics[width=0.8\columnwidth]{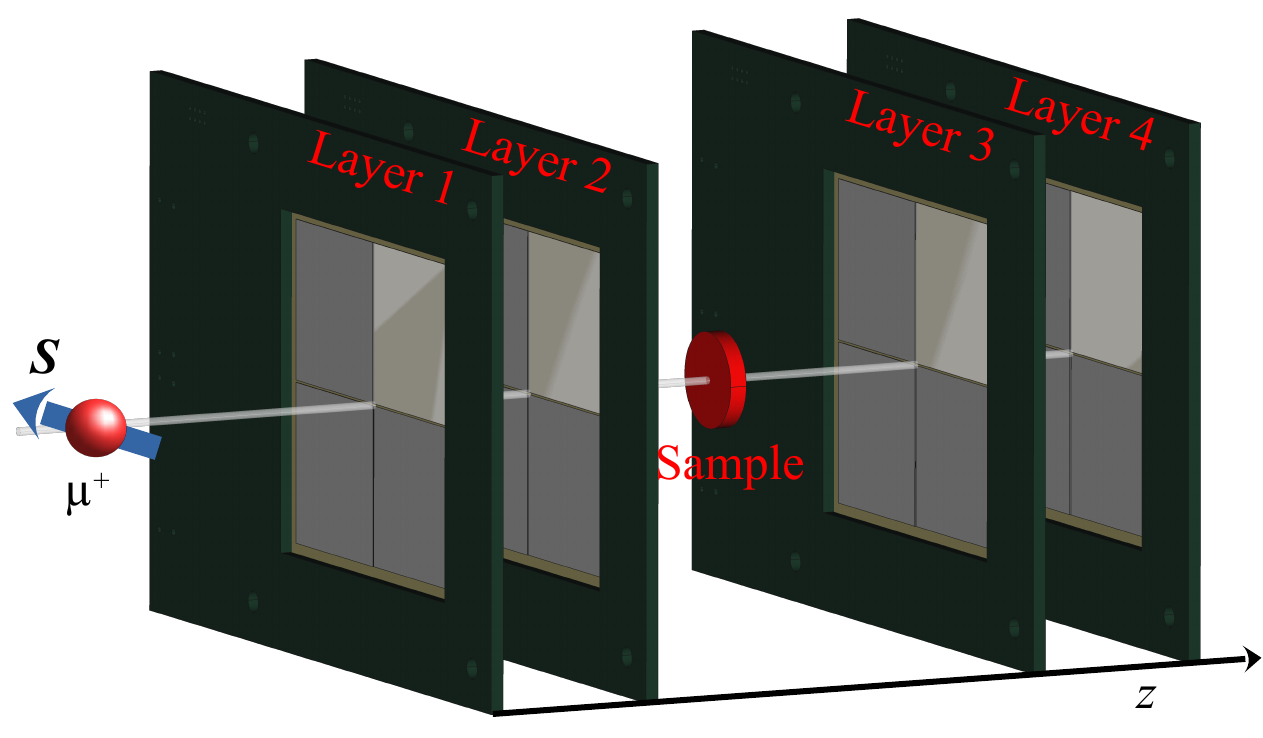}}
    \caption{A schematic of the spectrometer prototype which consists of four layers with the sample in the middle. The grey line indicates the beam propagation direction (along $z$) and the blue arrow denotes the initial muons spin direction, perpendicular to the beam.}
    \label{Spect}
\end{figure}

We have fabricated "Quad" modules consisting of four MuPix11 Si-pixel chips \cite{mupix10} mounted on a single printed circuit board (PCB, Fig.~\ref{spectrometer}(a)). The chips were glued onto a \SI{20}{\micro m} thick polyimide foil to provide mechanical stability. The thickness of the MuPix11 chips used in the modules varied between \SI{50}{\micro m} and \SI{100}{\micro m}, which was sufficient for a first prototype aimed at exploring the general capabilities of this technology (see Simulations section). For the reported measurements, we used four such Quads to construct a basic \mSR\ spectrometer, with two Quads forming the upstream detector set and two forming the downstream detector set (Figs.~\ref{spectrometer}(b) and~\ref{Spect}).
The measured samples were suspended on a frame using aluminized mylar tape and mounted in the middle between the two detector sets.
The design of this spectrometer allows for a flexible positioning of both Quads and sample relative to each other.
However, for the reported measurements in this work the Quads were placed at distances of $z=\pm 10$~mm and $z=\pm 30$~mm at either sides of the sample at $z=0$. 

The electronics and data acquisition (DAQ) software for the spectrometer were adapted from the Mu3e collaboration \cite{mu3edaq,ieeeDataFlow}. The Quad PCB was fed by a low voltage (LV, typically \SI{1.9}{V}) to operate the Si-pixels and a high voltage (HV, typically \SI{-30}{V}) to bias their substrate (Fig.~\ref{spectrometer}(a)).
The data was read using a custom made front-end board (FEB) \cite{mu3edaq,ieeeDataFlow}, converted to optical signals and then fed into the DAQ PC hosting the DE5a-NET development board~\cite{de5net}. All our measurements were performed with a muon spin polarization perpendicular to the beam direction, as indicated in Fig.~\ref{Spect}.

\section{Simulations} \label{Simulations}
\begin{figure}[h]
\centering
\includegraphics[width=0.8\columnwidth]{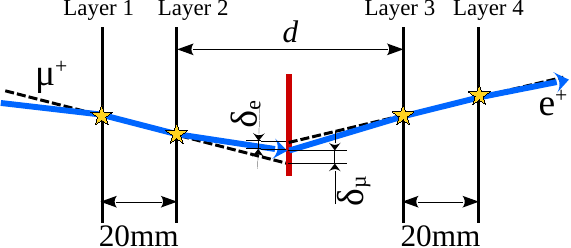}
\caption{The actual (blue solid lines) and reconstructed (dashed lines) trajectories of an incoming muon and an emitted positron as they pass through the Si-pixel chips. The red line represents the sample centered between layers 2 and 3.}
\label{Tracks}
\end{figure}

At the energy scales relevant for \mSR\ measurements (\SI{\sim 4.1}{MeV}), the primary source of uncertainty in the vertex reconstruction of an incoming muon trajectory, $\delta_\mu$, is predominantly due to multiple Coulomb scattering within the inner silicon chip, as illustrated in Fig.~\ref{Tracks}.
Such scattering introduces an uncertainty, $\delta_e$, in determining the trajectory of emitted positrons as well, however, to a lesser extent due to their relatively high kinetic energy. Therefore, $\delta_\mu$ can be considered as an upper limit on the resulting lateral resolution of the vertex reconstruction scheme. 

\begin{figure}[h]
\centering
\includegraphics[width=0.8\columnwidth]{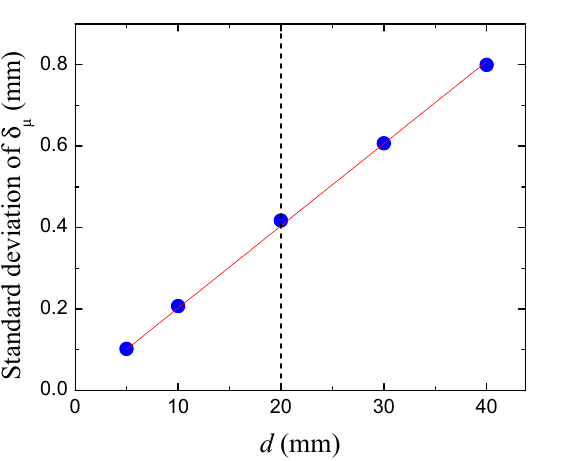}
\caption{The standard deviation of the distance between the actual and extrapolated muon position on the sample as a function of the distance between layers 2 and 3. The solid line is a linear fit of the data. The vertical dashed line marks the experimental conditions used here.}
\label{STDvsd}
\end{figure}

To estimate this uncertainty, we employ musrSim simulations \cite{musr-sim, musr-sim2}, based on GEANT4 \cite{geant4}, to model the scattering behavior of muons traversing two silicon layers of thickness \SI{50}{\micro m} (layers 1 and 2 in Fig.~\ref{Tracks}). We also take into consideration the polyimide foil (\SI{20}{\micro m}) onto which the chips are glued. The simulations provide the coordinates of the muon hits on each layer, $\mathbf{r}_i$, as well as the actual landing coordinates at the sample, $\mathbf{r}_s$. We use the hit coordinates in the upstream set to perform a linear extrapolation to the sample position, $\mathbf{r}_s^{ext}$. Finally, we calculate the distance between the actual and extrapolated muon position at the sample, i.e., $\delta_\mu = |\mathbf{r}_s - \mathbf{r}_s^{ext}|$ for all incoming muons. The standard deviation of $\delta_\mu$ was used as a measure of the uncertainty and evaluated from simulations with varying distance, $d$, between layers 2 and 3. As seen in Fig.~\ref{STDvsd}, the $\delta_\mu$ increases linearly with $d$ at a rate of \SI{\sim 0.02}{mm} per mm spacing. However, it remains below \SI{\sim 1}{mm} up to $d \simeq 40$~mm.
In our configuration with $d=\SI{20}{mm}$, this uncertainty is calculated to be less than \SI{0.5}{mm} (dashed line in Fig.~\ref{STDvsd}). Moreover, further reduction of uncertainty can be achieved by combining the extrapolated tracks of the muon and its matched positron.

Efficient track extrapolation for both muons and positrons depends critically on the ability to distinguish between them. From the simulations we note that the energy deposited in the Si layers strongly depends on the particle species. In particular, we find that muons deposit a significantly higher energy compared to positrons. Therefore, this opens the possibility of an effective method to distinguish between particle species. Since the measured time-over-threshold (ToT) as implemented on the MuPix chip is proportional to the deposited energy of the hit, one can use this parameter in the future to identify muon and positron hits.

%% Alternative Version: 
%Efficient track extrapolation for both muons and positrons critically depends on the ability to distinguish between them. This is possible due to their substantially different energy deposition in our energy regime. The time-over-threshold (ToT) measured on the MuPix chip is proportional to this deposited energy. In the future, this parameter could be used to identify muon and positron hits.

\section{Data Analysis}
The data obtained from the MuPix11 chips are a stream of detected hits with their lateral coordinates $(x,y)$ on the chip and the corresponding time-stamp $t$ of the hit. There are two types of valid hits, (i) an incoming muon which stops in the sample and (ii) an emitted positron from the sample.
Both must have coincidence hits in two layers, either the upstream set (layers 1 and 2) or the downstream set (layers 3 and 4). Any coincidence hits in more than two layers represents either a muon missing the sample or a positron that does not originate from a muon stopping in the sample, and are therefore ignored.

Using the Corryvreckan software framework \cite{corry,corry2}, the raw data of the recorded hits were filtered to obtain only such valid events. These were then used to construct tracklets, i.e., two points $\mathbf{r}^1_i=(x^1_i,y^1_i,z^1_i)$ and $\mathbf{r}^2_i=(x^2_i,y^2_i,z^2_i)$ representing the coordinates of the hit $i$ in layers 1 and 2 (or layers 3 and 4), respectively, both having the same time-stamp $t_i$. Here, the coordinate $z$ represents the position of the layer along the beam direction as show in Fig.~\ref{Spect}.

In order to generate the \mSR\ histograms, we used a simple algorithm, where we identify an incoming muon trajectory as a hit in layer 1 with coordinates $(x,y)$ that fall within a 4~mm diameter from the center of the beam collimator (a Pb cylindrical collimator with a 3~mm diameter hole). Using $\mathbf{r}^1_i$ and $\mathbf{r}^2_i$ of such hits, we extrapolate the trajectory to the sample position at $z^s_i=0$, $(x^s_i,y^s_i,0)$. Then, we open a software data gate of \SI{13}{\micro s}, and look for another trajectory within this time window that can be extrapolated to the vicinity of $(x^s_i,y^s_i,0)$. More specifically, within a distance $d_{\rm match}$ with a typical value of \SI{1}{mm}.
If a matching trajectory, $j$, is found, we record the time difference $t_j-t_i$ in the corresponding upstream or downstream histogram. Otherwise, we move on to the next muon trajectory candidate. Note that during the data analysis, for incoming muons rate of \SI{\sim 40}{kHz}, we only find very few events with two positron trajectories that match the same incoming muon candidate. This is to be expected since the probability of two muons landing within \SI{\sim 1}{mm} from each other is $\sim 0.6\%$ at this rate.
% Calculation from Hubertus
%At rate 40kHz there are 23kHz pile ups for 10us data gate (Poisson statistic)
%For a beam of diameter 4mm has an area of ~13mm^2.
%The probability that a pileup happens in the same 1mm^2 is 1/13 * 1/13 = 1/169. Total pileup rate in 1mm^2 at 40kHz is therefore
%RP = 23 kHz / 169 = 0.13 kHz
%relative: 0.13kHz/40kHz = 0.3%
The algorithm described above, although very simple and can be improved in the future, was sufficient to enable measurements of vertex reconstructed \mSR\ (VR-\mSR) signals accurately, reproducing measurements performed on the "conventional" GPS~\cite{gps} spectrometer at PSI as described below.

\section{Results}
\subsection{VR-\mSR\ Spectra}
\begin{figure}[h]
    \centering
    \includegraphics[width=\columnwidth]{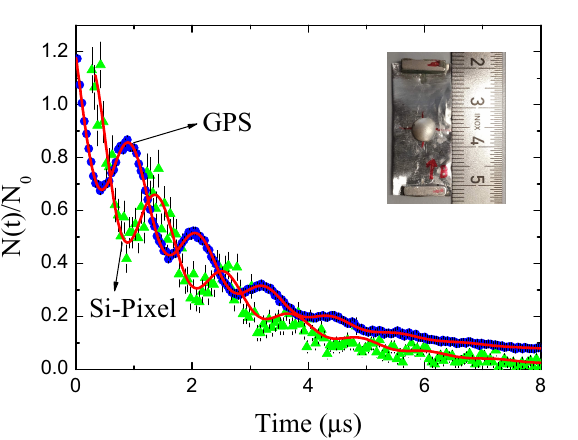}
    \caption{A first VR-\mSR\ signal measured from a \SI{6}{mm} diameter Al disc placed between two permanent magnets producing a transverse field of \SI{\sim 6.3}{mT} at the center (inset). The number of counts as a function of time measured in the upstream detector of the Si-pixel spectrometer (green triangles). This is compared to the counts in the upstream detector of GPS (blue circles) measured on the very same sample. The solid lines are fits as described in the main text.}
    \label{SipixelGPS}
\end{figure}

Since our prototype spectrometer does not have a magnet, we use permanent magnets placed above and below a \SI{6}{mm} diameter Al disc to generate a magnetic field which is perpendicular to the initial muon spin polarization (see inset of Fig.~\ref{SipixelGPS}).
Measurements of this sample using the GPS spectrometer as a reference show a clear precession signal in an applied field of \SI{\sim 6.3}{mT}, as seen in Fig.~\ref{SipixelGPS}.

\begin{figure*}[htb]
    \centering
    \includegraphics[width=\textwidth]{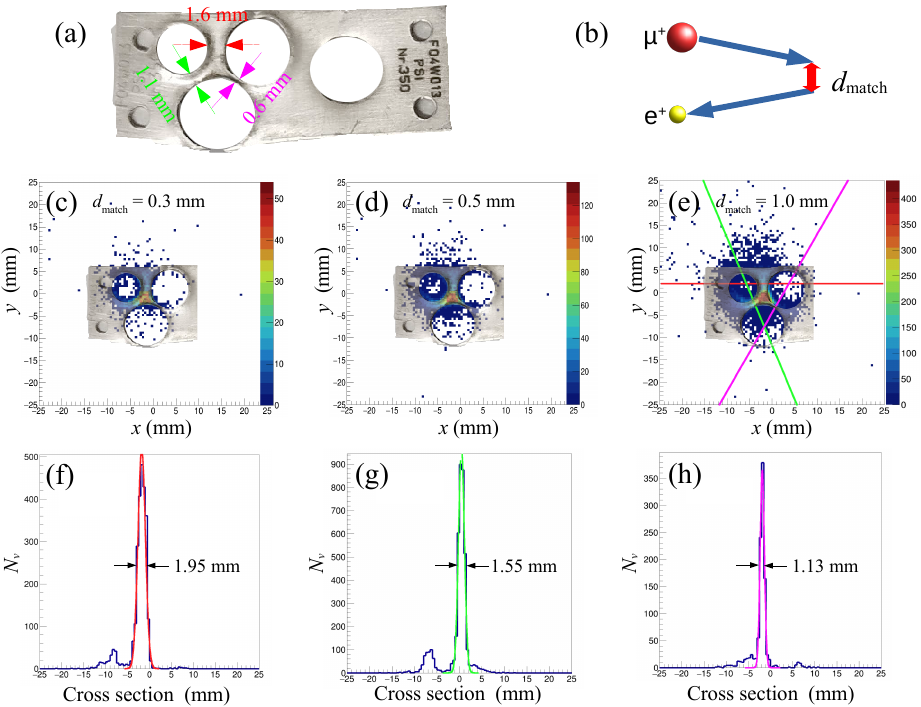}
    \caption{(a) The measured silver sample with indication of the typical sizes of its features. (b) A schematic showing the condition for matching an incoming muon to an emitted positron.  The two dimensional histogram of the coordinates of vertices of good muon-positron events at the sample position using a $d_{\rm match}$ of (c) \SI{0.3}{mm}, (d) \SI{0.5}{mm} and (e) \SI{1.0}{mm}. The sample image is scaled and overlaid on the plots showing that the vertices reflect the sample shape. The number of detected vertices, $N_v$, along the (f) red, (g) green and (h) magenta cross sections in (e). The solid lines in (f)-(h) are fits of the peaks in $N_v$ using a Gaussian curve with the corresponding FWHM.}
    \label{lateralres}
\end{figure*}
We detect the same precession signal (with a different phase due to the different geometries) in the GPS and the Si-pixel prototype spectrometers. Both spectra were fit (single histogram fits, solid lines) using,
\begin{equation}
    N(t)=N_0 e^{-t/\tau_\mu} \left(1+A\cos (\omega t+ \phi)e^{-(\sigma t)^2/2} \right) + B,
\end{equation}
where $N_0$ is the initial counts, $A$ is the asymmetry parameter, $\tau_\mu =$~\SI{2.197}{\micro\second} is the muon's lifetime, $\omega$ is the precession frequency, $\phi$ is the phase, $\sigma$ is the damping rate and $B$ is the accidental background counts. The results of these fits give an equal precession frequency (Si-pixel: \SI{0.85 \pm 0.01}{MHz}, GPS: \SI{0.855 \pm 0.001}{MHz}) and damping rate (Si-pixel: \SI{0.323 \pm 0.055}{1/\micro s}, GPS: \SI{0.285 \pm 0.007}{1/\micro s}) within the error bar.
It is important to point out here that the algorithm used to create the \mSR\ histograms from the Si-pixel spectrometer has (by design)  a negligible accidental background, $B$, compared to the currently used spectrometer technology (Fig.~\ref{SipixelGPS}).
This is because good events associate each muon with a single corresponding positron. Therefore, we conclude that vertex reconstruction schemes enable measurements with much longer data windows without reducing the data rate or compromising time resolution. We also note, that the signal in the Si-pixel detector is from a data set of $\sim35$ seconds duration only as opposed to the $\sim 14$ minutes measurement time at GPS. The limited run time on the Si-pixel spectrometer was due to DAQ software issues during the reported measurements which caused data corruption beyond that time.

\subsection{Lateral resolution at the sample}
In order to explore the lateral resolution limits from vertex reconstruction schemes on the sample size and geometric features, we inspect the lateral distribution of vertices of valid events.
For this purpose we use a silver plate with various cutouts as a sample (Fig.~\ref{lateralres}(a)). The vertices are defined as the intersection points (within a distance of $d_{\rm match}$, see Fig.~\ref{lateralres}(b)) of an incoming muon trajectory and a corresponding emitted positron trajectory at the sample position, $z=0$. We can clearly resolve details as small as \SI{\sim 0.6}{mm} on the sample as can be seen in Figs.~\ref{lateralres}~(c-e). 
Moreover, we vary the value of $d_{\rm match}$ to study the impact of our selection criterion. As an example, we show the vertices using $d_{\rm match}=0.3$~mm, \SI{0.5}{mm} and \SI{1.0}{mm} in Figs.~\ref{lateralres}(c), (d) and (e), respectively. In order to quantitatively evaluate the width of the features in Fig.~\ref{lateralres}(a), we plot the number of reconstructed vertices, $N_v$, along the corresponding cross section, as shown in Figs.~\ref{lateralres}~(f-h) for $d_{\rm match}=1.0$~mm. The peak across the corresponding sample's feature is then fit using a Gaussian function and used to evaluate the full width at half maximum (FWHM). As summarized in Table~\ref{tab:resFWHMCut}, we find that the FWHM values increase with the width of the feature, while overestimating it by $\sim 0.4$~mm. 
\begin{table}[h]
    \centering
    \renewcommand{\arraystretch}{1.2}
    \begin{tabular}{|c|c|l|l|l|}
         \cline{3-5}
        \multicolumn{2}{c|}{}  &  \multicolumn{3}{c|}{\textbf{$\mathbf{d_{\rm match} (mm)}$}} \\ \cline{3-5}
        \multicolumn{2}{c|}{}  & \ \ 0.3 \ \ & \ \ 0.5 \ \ & \ \ 1.0 \ \ \\ \hline
        \textbf{feature} & \ \ 1.6 \ \ & \ \ 2.19(10) \ \ & \ \ 1.98(5) \ \ & \ \ 1.95(3) \ \ \\ \cline{2-5}
        \textbf{width} & \ \ 1.1 \ \ & \ \ 1.60(7) \ \ & \ \ 1.62(5) \ \ & \ \ 1.62(2) \ \ \\ \cline{2-5}
        \textbf{(mm)} & \ \ 0.6 \ \ & \ \ 1.30(17) \ \ & \ \ 1.25(10) \ \ & \ \ 1.41(5) \ \ \\ \hline
    \end{tabular}
    \caption{The feature width (left column) for different $d_{\rm match}$ values (top row) compared to the FWHM of the vertices of good events along the corresponding cross section.}
    \label{tab:resFWHMCut}
\end{table}

However, the estimated width does not vary much regardless of $d_{\rm match}$. The main difference is the number of events which decreases significantly with decreasing $d_{\rm match}$. This indicates that, at least at an incoming muon rate of \SI{40}{kHz}, our algorithm manages to select good events very efficiently and accurately even for $d_{\rm match}~\sim~\SI{1}{mm}$. We note that for higher rates the $d_{\rm match}$ may need to be reduced to allow for a good separation of individual muons.

\section{Summary and Conclusions}
We have built a novel \mSR\ spectrometer prototype based on MuPix11 Si-pixel chips that provide information about the lateral position of the particle hits as well as their time. Using this information to obtain the landing position of an incoming muon on the sample and corresponding emitted positron, we are able to construct VR-\mSR\ histograms using vertex reconstruction schemes. With this added information compared to conventional \mSR\ we are able to perform measurements on samples of sub-millimeter size which are not possible using current spectrometer technology. In addition, this method opens the possibility of performing \mSR\ measurements on multiple small samples simultaneously or measuring a single large sample at rates more than 10 times higher than currently possible at continuous muon source facilities.

The presented prototype is operated in air, without a cryostat to host the sample, and without the possibility to apply an external magnetic field. Operation of the Si-pixel detectors in vacuum, to allow the introduction of a cryostat, is crucial for future utilization of this technology for standard \mSR\ measurements. In addition, to allow extrapolation of muons' and positrons' trajectories in a magnetic field, a three layer configuration may be needed for accurate vertex reconstruction. These issues are currently being addressed at PSI.

\section{Acknowledgments}
This research is funded by the Swiss National Science Foundation (SNF-Grant No. 200021\_215167).
All experiments were performed at the Swiss Muon Source SµS, Paul Scherrer Institute, Villigen, Switzerland. The authors are grateful to Frank Meier, Niklaus Berger, Andreas Knecht and Alex Amato for fruitful discussions. We also thank Martin Müller for his initial help porting the Mu3e DAQ system.

%\section{Author Contributions}

\bibliographystyle{apsrev}

\bibliography{main}

\end{document}